# Interact and React: Exploring Gender Patterns in Development and the Impact on Innovation and Robustness of a User Interface Tool

Dr Siân Brooke

*University of Amsterdam, Informatics Institute, Digital Interactions Lab*
s.j.m.brooke@uva.nl

In open-source software design, the inclusion of women is often highlighted simply to remind programmers that women exist. Yet, little attention is given to how greater gender diversity, specifically women's participation, could fundamentally alter development patterns. To understand the potential impact of gender inclusion, this study investigates React, a widely used JavaScript library for building user interfaces with an active contributor community. I examine gender differences in metrics of robustness and innovation, as well as shifts in contribution patterns leading up to major version releases over 11 years of the React project. My results show that the exclusion of women is detrimental to software as women contribute significantly more to feature enhancement and dependency management. By exploring how gender influences innovation and robustness in the development of React, the study offers critical insights into how increasing gender diversity could lead to more inclusive, innovative, and robust software.

CCS CONCEPTS • Human-centered computing → Human computer interaction (HCI); Empirical studies in HCI.

**Additional Keywords and Phrases:** Open-source software; Gender; Design; GitHub; Collaboration

## 1 INTRODUCTION

Hand tools, like hammers and screwdrivers, are often designed with the assumption that men are the primary users, resulting in features such as larger grip sizes and ergonomics that are less efficient and comfortable for women. This gendered approach to tool design reflects broader societal assumptions about who performs certain crafts, often marginalizing women before we even consider what the tools might create [14]. Similarly, while valuable work has been done in Human-Computer Interaction (HCI) to examine bias in interfaces [18, 33] and algorithms [26], much less attention has been given to the early stages of creation—such as programming languages and the tools used to build software. These foundational components, like hand tools, are where gender bias is encoded into software through women's exclusion, raising the question of how the inclusion of more women might influence the development of technology.

Recognition of how biases can be embedded in software has increased significantly in recent years [33]. Some biases are obscured in black-box algorithmic structures [10] whilst others are visible to the user in the interface itself [4,

33]. For example, sorting algorithms on freelancing platforms have been found to discourage the hiring of women and non-white users by reinforcing biases through social feedback mechanisms such as reviews [10]. Layered on top of algorithms, the design of such interfaces has the power to create belonging [18] or to exclude and discourage participation [7]. Nonetheless, software and interfaces are not independent technical artifacts. Rather, a select few tools written in JavaScript (.js) are responsible for an estimated 75% of user-interfaces, namely, React (~40%), Vue (~20%), and Angular (~15%). [1,2] JavaScript is the most popular programming language, used by 68% of developers; but women are heavily marginalized: in 2023, they represented only 6% of Javascript users

Gender representation is a central focus of HCI research into software biases. Valuable research in the field has shown how gender groups often experience and interact with software in varied ways, leading to potential disparities in usability and inclusion [8, 17, 18]. For instance, gender biases are reflected and amplified in search engines, reinforcing stereotypes [12]. Criado-Perez [5] illustrates this by showing how a search for "computer programmer" often favors male programmers' websites, as algorithms tend to associate masculine pronouns more closely with the term. If software design overlooks these gender biases platforms can reinforce existing inequalities [25]. Understanding how gender influences the development of user-interface (UI) tools is crucial, as it ensures that these tools are designed to be inclusive, fostering environments where all users feel valued and empowered to participate fully.

HCI contributed greatly to understanding exclusion and working towards inclusion. How participation to open source software (OSS) projects is valued and measured [17] has implications for understanding how metrics of success are biased in favor of men [1]. CHI scholarship has examined the impact of gender in software development in terms of women exclusion or patterns of contribution to OSS more broadly [17, 30]. There is also a strong body of work on identification of biases in interfaces [4] or computational techniques [26] and empirically substantiates remedies for gender-based exclusion [6, 33]. Not sufficiently considered by previous studies is how gender impacts key moments or releases in the software development process. Previous work either flattens contributions and analyses the totality of a user's participation or analyzing the end product once the software or interface is already in use.

By not considering *time* in the design of software it is difficult to ascertain how the exclusion of women may impact the process at crucial moments. The prevailing logic is that excluding women from software development results in biased tools that mirror the perspectives of their designers who are predominantly men [7]. However, women's contributions to development and design go far beyond their identity. When women engage in technical projects, they offer much more than just a reminder of their presence and existence to programmers who are men. This leads to the question; *what do we lose by excluding women from software development?*

To address this question, I explore on the case study of React, an UI component library created using JavaScript designed specifically for creating interfaces. Created by Facebook (Meta) in 2011 to support its News Feed functionality, React was then open sourced in 2013 becoming heavily supported by external contributions through the platform GitHub. React represents an OSS project with a significant history in the development of UI technologies.

As a first step, I defined key indicators of quality in software development focusing on *robustness* and *innovation* over time. *Robustness* is defined as the ability of software to handle errors and unexpected conditions, including bug fixes and clear usage documentation. *Innovation* then involves introducing new features or methods enhance software performance and user experience, often marked by significant version releases. While challenging bias is crucial, unpacking gender differences in innovation and robustness of contributions will also help us understand how greater inclusion can transform the software development process—something that is often assumed but rarely examined in depth. To explore the relationship between gender and software development, this study poses two key research questions: *(1) Does gender influence the robustness of development of React? (2) Does gender impact innovation in the development of React?*

---

[1] State of JavaScript 2023: https://2023.stateofjs.com/en-US
[2] Stack Overflow Developers Survey 2024: https://survey.stackoverflow.co/2024/

To address these questions, I examined the large-scale and freely available dataset on GitHub Archive, filtering for activity under the "Facebook/React" organization. The archive was supplemented with additional user information required for gender inference using the GitHub REST API, which allows for programmatic access to GitHub data, such as repository activity, user profiles, and contributions. Given the importance of work on inclusion software development the analysis in this paper prioritized interpretability over complexity, ensuring that the findings were clear and accessible rather than obscured by overly intricate methods. The contributions are:

- Identifying gender bias and the need for intervention within the first 20 months of contribution.
- Introducing a temporal analysis of gendered contribution patterns by focusing on version releases.
- Showing how the exclusion of women impacts the robustness and innovation of OSS.

Beyond the study presented in this paper, I have also provided supplementary material, including the dataset used in the study in the following repository: [*Figshare data repository DOI*]

## 2 RELATED WORK

### 2.1 Gender and Software Design

There is plentiful evidence that the dominance of men in software engineering has consistently encoded bias into the design of interfaces. Consequently, the underrepresentation of women in computing has fostered an association between masculinity and technical expertise, perpetuating the idea that men are "natural" programmers [14]. This masculine control over technical fields further shapes the very definition of success and legitimate contributions, embedded in programming culture from individuals first interactions.

The dominance of men in programming shapes the perception that users are "male-by-default", which significantly impacts how software is designed and used [2, 30]. Since there are gender differences in how software is utilized, this bias results in software being tailored to men's needs and preferences. As a result, women are marginalized not only in software development but also in its usage [22, 33] Research across various fields—such as spreadsheets, visualization systems, online education platforms, home automation, intelligent agents, and programming tools [33]—demonstrates that software often favors male-dominated usage patterns due to men's predominance in development.

In looking to research on the design process, Williams [34] showed how gender bias can be embedded at any point in the product cycle, as women's perspectives were rarely incorporated into engineering practices. Moreover, the "Gender Inclusiveness Magnifier" (GenderMag) marked a significant shift by introducing a method to identify gender biases in how user interfaces are designed [4]. It specifically focuses on uncovering biases in the "ways of thinking" (cognitive processes) and workflows that these interfaces support [4]. Extending this work, Metaxa-Kakavouli et al. [18] find that the way interfaces are designed impacts ambient belonging, or the sense of fitting in with a community or culture. For example, interfaces designed with masculine themes, such as Star Trek imagery or styles reminiscent of a computer terminal, discouraged women from engaging with the platform [18]. The findings supported the notion that web interface design can activate gender biases, underscoring the importance of creating more inclusive user interfaces. [18]. Going a step further, Vorvoreanu et al [33] show how the HCI method to detect bias in GenderMag can generate more inclusive designs, finding that designing for cognitive diversity more generally improves software's gender inclusiveness.

Considering this evidence, examining how women's contributions to software development evolve over time is essential for understanding and addressing gender bias in the field. While tools like GenderMag have been instrumental in identifying biases in interface design, the next step is to explore how increasing the participation of women in software creation can alter the development processes. This understanding is crucial if we are to move beyond mitigating bias and towards fostering truly inclusive, equitable development environments.

**2.2 Defining Quality in Contributing**

Understanding how "quality" and "success" are defined in computing is crucial for studying how gender affects the development process, as these definitions often reflect biases that can influence who is recognized and valued in technical contributions. Fundamentally, the association of programming expertise with masculinity leads to success in OSS being linked to practices favored by men [16]. Riley [23] highlights how certain linguistic choices in defining quality within technical disciplines are inherently exclusionary. She argues that this language marginalizes alternative ways of knowing and reinforces the dominance of epistemologies aligned with the perspectives of heterosexual, white, men [23]. Terms like "rigor", with their connotations of hardness and inflexibility, are closely tied to masculinity and often devalue work that emphasizes creativity, flexibility, or relational approaches—qualities more commonly associated with femininity [23].

Knowledge of a programmer's gender can influence the language that is used to ascribe quality to their work. For instance, when women do write code, their gender significantly influences how its quality is perceived, with their work often being described as "neat," "elegant," and "pretty" rather than "strong" and "rigorous" [15]. Brooke's [3] analysis of Python contributions on GitHub shows that while there are stylistic differences between code written by men and women, the actual quality of the code is the same. Therefore, gendered perceptions of quality in software development are shaped by biased language, which reinforcing masculine ideals while diminishing the value of women's contributions

Taking a more granular approach to technical contributions, the association of masculinity with quality diminishes the value of the tangible contributions women make to technical projects. Aligning with Riley's [19] conclusions, writing code to create new features is often regarded as inherently more valuable and even morally superior to non-code contributions [2, 19]. Tasks like organization and community building, more frequently undertaken and appreciated by women, are often devalued because they fall outside the male-dominated mainstream of coding work [19].

In contrast to women, men's contributions are typically viewed as the "real" work of coding, aligned with values like innovation and speed [19]. Critical research emphasizes the importance of community-centric roles, which, despite being crucial to project success, are often invisible in contribution metrics [2]. Moreover, computational studies reveal that gender-balanced teams tend to be more productive and experience lower turnover [32]. Thus, gender bias not only marginalizes diverse contributions essential to project success but also reinforces gendered hierarchies and the association of masculinity with "success" in open-source development.

Building on this literature, I define contributions to OSS in terms of *robustness* and *innovation*. Robustness emphasizes adaptability and flexibility, qualities less tied to gendered notions than "rigor" [23]. Although innovation literature has been critiqued for its gender blindness, Pecis [21] finds that when women's contributions to innovation are recognized, they are respected and valued. Thus, innovation is better understood as an interactive process based on collaboration and knowledge-sharing, rather than a competitive space that inherently excludes women.

### 2.3 Collaboration in Open-Source

Narrowing our scope to collaboration platforms, research on gender differences in programming forums has often focused on "gender gaps" in individual user activity (i.e., 1, 14, 28) . Scholarship has highlighted how narratives of women's exclusion can conflate the platforms' narrow definitions of participation with broader concepts of quality and success in programming as a whole [1, 31]. For example, May et al. [16] observe that women are less successful on Stack Overflow in terms of a "reputation score", due to gender differences in the level and type of activity. They find that men answer more questions than women but they also receive more rewards for their answers, even after accounting for confounding factors like tenure [16]. Similarly, Vedres and Vasarhelyi [31] demonstrate that typical interactive behavior patterns of women on GitHub are not rewarded, resulting in a lower probability of continued participation. Sultana et al [28] also uncovered significant gender biases in OSS projects on GitHub, particularly in code reviews where women faced longer review times and lower response rates compared to men.

Studies also examine women's agency is strategies to overcome gender discrimination in collaborative environments, such as choosing to remain anonymous. Significantly, Terrel et al. [29] found that women's contributions to OSS on GitHub—whether to code, documentation, or other project resources—are more likely to be accepted than men's when their gender is obscured. However, when their gender is visible, women's contributions are 15% less likely to be accepted. In fact, they found that women's contributions were especially likely to be accepted to files in JavaScript, Ruby, and Go [29]. Nonetheless, more recent work has also found that anonymity provides little advantage for women, as it is the interactive and collaborative behavior that disadvantages them [31]. Additionally, anonymity may lead to a lack of trust and therefore exclusion from OSS projects [28, 31].

Building on research that examines gender in design, collaboration, and definitions of quality in software development, I explore whether greater inclusion of women—or reducing their exclusion—could reshape the development process. By exploring how women's increased participation might influence the creation of software tools, the study aims to assess whether a more diverse and gender-balanced development environment impacts the robustness and innovation of technical tools, using React on GitHub as a case study.

### 2.4 Introducing GitHub and React

GitHub is a remote collaboration platform that allows users to work together or independently on technical projects. Used by 93% of developers in 2023,[3] It can track changes made to files, launch software, and host websites through projects knows as "repositories" (or *repos*). GitHub has remained at the forefront of OSS and technological development, hosting Interactive Developer Environments (IDEs), including the most popular source code editor, Microsoft's VS Code.

GitHub is fundamentally interactive. It fosters collaboration and continuous improvement of software through its core features; *commits* and *pull requests*. Each "commit" to a project represents a snapshot of code changes, allowing users to document and share their progress incrementally. Commits generally fix something specific, such as correcting a typo in the description of block of code. This commit history not only tracks the evolution of a project but also enables others to see the rationale behind each change through the text of "commit messages".

Another particularly pivotal feature is "pull requests" (PRs), enabling developers to propose changes to a codebase (repository) and engage in discussions about those changes before they are merged with the current version of a project. This process is also inherently interactive, as it encourages peer review and feedback from multiple contributors, ensuring that code quality is maintained and even improved over time. Through these features, GitHub transforms software development into a dynamic, collaborative process, where code is continuously refined and enhanced through collective effort [31].

---

[3] https://github.blog/news-insights/research/the-state-of-open-source-and-ai/

Although GitHub enables broad participation in OSS, it is not a neutral platform [20]. Designed primarily by and for male programmers, it reflects and amplifies the gender biases inherent in OSS [17]. Features like pull requests, often seen as assertive, are more frequently used by men, reinforcing male-dominated behaviors and contributing to the exclusion of women [20]. However, despite evidence of these biases, studying gender dynamics on GitHub remains essential due to its central role in OSS development. GitHub provides a vast dataset for understanding participation and success, and examining its biases can reveal broader patterns that affect inclusivity across other digital platforms [26]. While it is a valuable resource for researchers, its male-centric design influences how success is defined and how gender and diversity shape team productivity [30]. Addressing these biases is key to promoting more inclusive development environments in the OSS community.

Hosted and maintained on GitHub, React is a tool (specifically a JavaScript library) that helps developers build UIs for websites, making it easier to create interactive and dynamic web pages. React was originally created by Facebook specifically to improve the performance and manageability of their News Feed feature. In 2011, Jordan Walke, a Facebook engineer, developed React to handle the complex and dynamic user interface requirements of the News Feed. React proved successful internally and was made publicly available on GitHub by Facebook in 2013, allowing developers worldwide to use and contribute to it. Despite being founded over 11 years ago, the 2023 *State of React*[4] survey found that 92% of React Developers identify as a Man, 6% as a Woman, and 1% as Non-Binary or Gender Non-Conforming. React epitomizes the dominance of men in open-source projects on GitHub.

React has risen to immense popularity in the software development community due to its innovative approach and robust features. It allows developers to break down their website's interface into small, reusable pieces called "components", making it easier to build and maintain complex user interfaces. One of React's most popular features is the "virtual DOM", which improves performance by updating only the parts of the web page that have changed, rather than reloading the entire page. Another popular feature are "Hooks", which let you control and manage parts of your website's behavior, without needing to understand all the complex details of how everything works underneath. Because of such features React is considered very flexible; it can be used alongside other libraries and frameworks, fitting into different projects without requiring a complete rewrite.

Like GitHub, React is a tool frequently used by HCI researchers. HCI scholars have used React to design UIs, test new tools, and produce interactive visualizations [18, 27]. Furthermore, cutting-edge studies on interacting with Large Language Models (LLMs) also feature React. Not only is React used to build popular platforms like Instagram, Airbnb, and Netflix, but it also is utilized by chat-based LLMs in their user interfaces. Some notable examples include ChatGPT by OpenAI and Microsoft's Azure OpenAI. This has increased the use of React by researchers. For example, in their study of emotional bonds with conversational agents, Ha et al. [9] use React to create bespoke agent personas using the tool to allow study participant to interactively customize traits. There is considerable momentum in the study of LLMs to understand biases in the building and training of models, and the UIs we interact with should not be excluded from this effort.

---

[4] https://2023.stateofreact.com/en-US

## 3 METHODOLOGY

Figure 1: Construction of the Dataset

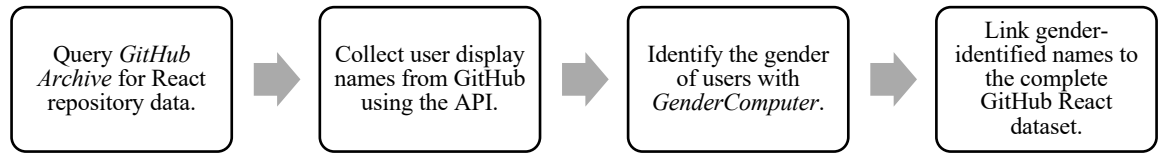

### 3.1 Dataset

This study is based on the GitHub Archive data, supplemented with the GitHub REST API to retrieve additional user information. The GitHub Archive[5] is a project created by Ilya Grigorik to store public events from GitHub, since February 2011. Currently, the system extracts data from GitHub Events API, where an "event" is any specific activity or action that occurs on GitHub, such as a push to a repository, the creation of an issue, or the submission of a PR, and it provides a record of that action along with relevant data. The data that is extracted is available in hourly archives that can be acessed using Google Big Query. These tables have fields with basic information about the repository, the actor of the event, and a JSON object for the detailed data known as the "payload".

As illustrated in Figure 1, I began by querying the GitHub Archive for React data and extracting the unique usernames from the dataset. Next, I used the GitHub REST API to retrieve profile information for these contributors, including their display name, location, email, and biography. Since the archive primarily focuses on events and provides limited user-level information, obtaining this additional profile data was essential for the gender inference process which is specified in section 3.3.

### 3.2 React Project

The React dataset for this study was compiled on 24th July 2024, with the first event on the GitHub React repository dating back to 26th May 2013 – when the project was open-sourced. Initially, I identified 1,246,593 events and 433,187 unique contributors to the repository. This study includes several React-focused repositories under the broader "Facebook/React" organization; React Website, React.js and React Native. React.js (or simply "React") is used for building web applications, focusing on creating UIs rendered in the browser, while React Native is used for building mobile applications on iOS and Android, using native mobile UI elements.

When analyzing difference in activity over time I focus on contributions leading up to releases. I include information on specific versions of React, starting from React 15.0.0, which introduced a new versioning schema and significant software changes. Each release is identified by a version number, typically following Semantic Versioning (SemVer), where, for example, version 16.8.1 represents a major version (16), a minor version (8), and a patch version (1). Whilst introducing significant enough changes to be included in this study, React 16.8.0 was not considered a major release due to SemVer as the new feature it introduced was backwards compatible, meaning that developers did not have to modify their existing codebase to accommodate the new version. In the final element of this study, I will examine if patterns of contributions by gender change as React approaches significant releases. The key milestones that will be analyzed are:

---

[5] https://www.gharchive.org/

(1) **React 15.0.0** (7th April 2016): Introduced significant changes in how React renders and updates the DOM, moving towards more efficient rendering.
(2) **React 16.0.0** (26th September 2017): Introduced Fiber, a new reconciliation algorithm, error boundaries, portals, and fragments.
(3) **React 16.8.0** (6th February 2019): Introduced Hooks, which allow using state and other React features without writing a class.
(4) **React 17.0.0** (20th October 20, 2020): Focused on making upgrades easier with no new features, aimed at improving the underlying architecture.
(5) **React 18.0.0** (26th March 2022): Introduced concurrent features, new hooks, and enhanced server-side rendering APIs.

The React data was then processed. First, I used dictionary matching to categories the PRs and commits by purpose by analyzing the text in the payload of the events. This typology was developed in reference to the Guide to Contributing,[6] updated through examination of the data, and refined and verified in discussions with three React Developers. As shown in Table 1, the typology of purpose was bug fixing, feature enhancements, documentation, refactoring, dependencies, testing. I also identified when another contributor was tagged in the body of a PR, as this can provide insights into collaborative dynamics and the distribution of influence within the development process.

Table 1: Contribution Purpose Typology

| Purpose Type | Example Terms | Description | Count |
|---|---|---|---|
| Bug Fixing | *Issue; Bug; Fix; Tagging a specific issue.* | Identifying and resolving errors or issues in the code. | 61,545 |
| Feature Enhancements | *Introduce; Launch; Sync* | Adding new functionality or improving existing features in the codebase. | 63,675 |
| Documentation Update | *Doc; readme; .md* | Revising or adding to the project's instructions to clarify usage, describe new features, or correct existing information. | 54,704 |
| Refactoring | *Reformat; Move; Refactor* | Structuring existing code to improve its readability, maintainability, or performance without changing its external behavior. | 60,217 |
| Dependencies | *Dependency; Version; Outdated* | Refer to external libraries, frameworks, or modules that a project relies on to function properly, typically documented in configuration files and managed through package managers. | 36,505 |
| Testing | *Test; Testing; Do not merge* | Verifying that the code behaves as expected by running automated or manual tests. | 35,007 |

---

[6] https://reactnative.dev/contributing/overview

The GitHub dataset offers an insightful case study of broader software development behavior, as it captures the dynamics of both open-source contributions and core contributions from Meta. React's development reflects the collaborative nature of OSS projects, with external contributors enhancing the codebase alongside Meta's internal team. This blend of community-driven and organizational contributions provides a comprehensive view of how large-scale, widely used projects evolve, showcasing patterns in collaboration and innovation that are relevant across the software development landscape more broadly. The React dataset created in this project is publicly available, including gender identification as discussed below.

**3.3 Gender Inference Procedure**

Gender inference was conducted using data from users' usernames, display names, locations, email addresses, and biographies. Although GitHub users log in with their usernames, their display names are often identical or similar to their given names, making them more effective for gender identification. This information was combined with names extracted from email prefixes, where possible, such as "jane rivers" from the email address jane.rivers@gmail.com. All these methods of gender inference were name-based, utilizing the Python tool *genderComputer*. Developed by Lin and Serebrenik [13], *genderComputer* was originally designed to infer the gender of users on the question and answer programming forum Stack Overflow and has been extended for use with GitHub [3].

*genderComputer* also incorporates location data to enhance the accuracy of gender inference. In the React dataset, 47% of users had their locations on their profile, with the 10 most popular countries being: China, United States, India, United Kingdom, Germany, France, Brazil, Japan, Canada, and Singapore. Representative of the wider GitHub community, the geographical spread highlights the global nature of GitHub's user base.

I supplemented the name-based gender inference with explicit self-declarations of gender, such as users stating their pronouns. I looked at the "about me" or biography of GitHub users' profiles, focusing on the use of gendered words and emojis. First, I matched specific gendered terms (such as "father") and pronouns (such as "she/her"). Second, I matched gender specific emojis that users may use to communicate their gender more subtly. This was aided by GitHub markdown which supports the inclusion of emojis. For instance, the emoji 👩‍💻 would be labelled as a "woman" as it depicts a "female technologist" and 👨‍🏫 would be labelled as "man" as it depicts a "male teacher".

Informed by Keyes [11], I intend to understand gender in a nuanced manner, with an analysis that includes the inference of gender beyond a binary. As my focus is on gender bias and discrimination, I use language of "man" and "woman". As an additional step towards gender inclusivity, if the user biography indicated a gender beyond a binary identification the user was labelled as "non-binary". For example, if they included the pronouns "they/them" or emojis such as ⚧ or a string of hearts in the non-binary flag colors. Given this explicit identification, if indications of non-binary identity were found in the user biography, this would override gender that was inferred from the login and display names or emails as it expresses agency in an explicit identification. In total, I inferred 975 non-binary contributors to React on GitHub.

It is worth noting that I found less woman users than non-binary users through the biography-based inference, with a total of 566 non-binary users actively declaring their gender on their profile. This may be because, while using their legal name in a display or login name can help women build a professional identity and gain recognition, openly identifying as a woman on their profile offers no clear advantage and could potentially be detrimental in exposing them to sexism. Additionally, since non-binary identities are more difficult to express through names alone, the biography section provides an important opportunity for these users to explicitly identify themselves.

Whilst I made efforts to include non-binary users in my analysis, not all the statistical tests chosen for simplicity and clarity supported the inclusion of such a small group. Furthermore, users without clear gender identification (woman, man, non-binary) were categorized as "anonymous". It was crucial to ensure that explicitly non-binary users were not conflated with anonymous users, as non-binary is a distinct and recognized gender identity category. This also allowed me to consider all contributors to React, and not reduce my sample to usernames suitable for gender inference. Balancing statistical interpretability with fair representation is crucial in computational inquiry. It is also essential to ensure that this approach does not inadvertently reinforce the gender binary through gender inference.

The results of the gender inference procedure for the contributors and contributions are outlined in Table 2. As well as the unique users, the table also includes contributions, interactions, and passive engagement. Contributions included activities such as identifying and fixing bugs, as well as making changes to code and documentation through commits and pull requests. Interactions are the comments on contributions such as comments on PRs and issues. Passive engagement, such as watching the repository is also included. Whilst this variety of activity is included in the dataset produced in this paper, the analysis focuses on contributions as this is directly relevant to software development.

Table 2: Result of Gender Inference Procedure

| Inferred Gender | Users Count | Users Percentage | Contributions Count | Contributions Percentage | Interactions Count | Interactions Percentage | Passive Count | Passive Percentage |
|---|---|---|---|---|---|---|---|---|
| Woman | 38,602 | 8.91% | 24,962 | 7.19% | 26,517 | 5.60% | 37,735 | 8.85% |
| Man | 198,347 | 45.79% | 179,105 | 51.60% | 255,612 | 54.0% | 198,431 | 46.56% |
| Non-Binary | 975 | 0.23% | 6,052 | 1.74% | 6,561 | 1.39% | 1,044 | 0.25% |
| Anonymous | 195,254 | 45.07% | 127,722 | 36.80% | 155,921 | 32.94% | 188967 | 44.43% |
| Bots | 9 | 0.0% | 9,236 | 2.66% | 28,728 | 6.07% | 0 | 0.0% |
| Total | 433,187 | | 347,077 | | 473,339 | | 426,177 | |

### 3.4 Operationalizing Robustness and Innovation

As discussed in the literature review, the gendered assumptions and weighted nature of language in software development influenced my approach to characterizing quality in this study. I specifically chose to quantify metrics related to *innovation* and *robustness*. These categories are informed by the gendered differentiation of activities that was highlighted in the review of HCI literature [19, 28, 31], ensuring a more inclusive and comprehensive analysis of contributions in OSS. Table 3 then details the specific indicators used in this study, along with the metrics and GitHub events that serve as the basis for each indicator.

Table 3: Indicators of Robustness and Innovation in Open-Source Software Development

| Development Indicator | Description | Event Type | Metric |
|---|---|---|---|
| Robustness | Code Review Practices | Pull Requests, Commits | (1) Merge rate <br> (2) Time to review a contribution |
| | Resolving Bugs and Issues | Issues | (1) Raising, resolving, and reopening issues <br> (2) Time to update or close an issue |
| | Contributor Retention Rate | All | (1) Time remaining active after first contribution. |
| Innovation | Diversity of Contribution Types | Pull Requests, Commits | (1) Proportion of contribution purpose |
| | Contributions Approaching Release | Pull Requests, Commits | (1) Patterns of contribution approaching version release |

*Robustness* encompasses activities related to organization and documentation, which are fundamental to open-source projects but are often devalued due to their associations with femininity [19]. In the context of OSS development robustness refers to the ability of a software system to cope with errors during execution and handle erroneous input gracefully. It implies that the software can function correctly under unexpected conditions and maintain its functionality without crashing or producing incorrect results. Robustness can be encapsulated in indicators such as the speed to which bugs or errors are fixed or even how well the purpose and function of the code is documented.

*Innovation* captures the typically masculine-associated contributions of new features and technical advancements [19] whilst maintaining an emphasis on interaction . Innovation involves introducing new ideas, methodologies, technologies, or features that significantly enhance the software's functionality, performance, user experience, or development process. Software version releases are pivotal to this innovation, serving as crucial milestones that encapsulate new features. These releases provide a stable foundation for further advancements, inspire new use cases, energize the developer community, and set benchmarks for future development[24]. By continuously improving the software and encouraging diverse contributions, version releases create an environment where innovation thrives, ensuring the software remains at the forefront of technological progress.

With the indicators established, the following section presents the results of my analysis of the impact of gender on robustness and innovation in the development process. All tests were bootstrapped due to the composition and size of the sample, ensuring more reliable estimates and accommodating potential variability in the data.

## 4 RESULTS

### 4.1 Robustness

The first research question asks: *Does gender affect the robustness of React's development?* To explore this, I examine key measures of participation and contribution over time, including the duration of the peer review process, the time taken to resolve bugs, and retention rates, all analyzed by gender. These measures are appropriate for assessing whether gender influences the efficiency and stability of the development process, with a focus on whether the time to complete these tasks differs based on gender.

*4.1.1 Peer Review Processes*

Figure 2: Distribution of Days to Merge or Close a Pull Request by Gender

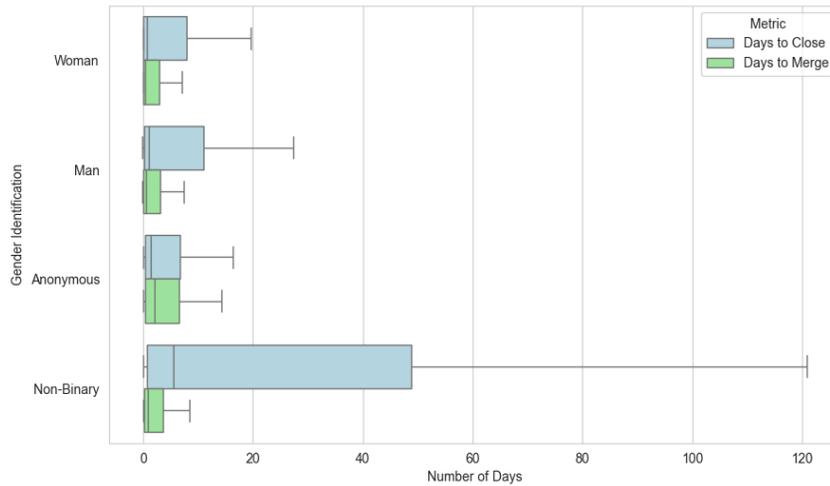

First, I examined the time it takes for PRs and commits to be resolved by gender, displayed in Figure 2. Merging a PR incorporates changes into the codebase, while closing it ends the PR without merging; merging often takes longer due to reviews and revisions. The findings reveal that women's contributions are merged the quickest, with an average (mean) of 7.09 days to merge and 27.57 days to close a PR. Men follow closely with 7.51 days to merge and 38.33 days to close. Contributions from anonymous users take slightly longer, averaging 9.14 days to merge and 16.58 days to close. Non-binary contributors experience the longest times, with 9.80 days to merge and 68.94 days to close. However, the differences in time to merge across gender groups was not statistically significant.

The time for a PR to be closed significantly differed (Welch's ANOVA=-4.64, $p<0.001$, df=2278), Games-Howell test shows that PRs from anonymous contributors are closed significantly faster than those from both men and women, with mean differences of 21.75 ($p<0.001$) and 10.99 days ($p<0.001$), respectively. Additionally, women's PRs are closed significantly faster than men's, with a mean difference of 10.76 days ($p<0.001$). Non-binary users were excluded from the post-hoc analysis due to the small sample size, which limited the ability to draw statistically meaningful conclusions. These results highlight significant differences in PR closure times based on contributor identity.

I also examined the merge rate by gender groups. On GitHub the *merge rate* refers to the proportion of PRs that are successfully merged into the main branch of a repository. Men have the highest merge rate at 24.62% follow by non-binary 21.73% and Women 20.38%. Anonymous contributors have the lowest merge rate at 3.98%. These differences were statistically significant ($p<0.001$). This shows that there are gender differences in the incorporation of contributions to the React library.

*4.1.2 Resolving Bugs and Issues*

I examined who is responsible for raising and resolving issues in the React library. Table 4 reveals distinct patterns in issue management based on inferred gender. Women have the highest percentage of issues raised, with 64.69% of their contributions falling into this category, but a lower percentage of issues resolved (33.04%). Men, while contributing the most in absolute numbers, show a more balanced distribution with 48.95% of issues raised and 48.77% resolved. Non-binary contributors have the highest resolution rate, with 89.46% of their contributions being resolutions, despite raising only 8.46% of the issues. Anonymous contributors have a high percentage of issues raised at 62.55%, but their resolution rate is significantly lower at 36.22%. Across all genders, the percentage of reopened issues remains low, hovering around 2-2.3%, with men having the highest percentage of reopened issues at 2.28%.

Table 4: Issues Raised, Resolved, and Reopened by Gender

| Inferred Gender | Raised | | Resolved | | Reopened | |
| --- | --- | --- | --- | --- | --- | --- |
| | Count | Percentage | Count | Percentage | Count | Percentage |
| Woman | 3,140 | 64.69% | 1,604 | 33.04% | 110 | 2.27% |
| Man | 22,081 | 48.95% | 21,999 | 48.77% | 1,027 | 2.28% |
| Non-Binary | 126 | 8.46% | 1,333 | 89.46% | 31 | 2.08% |
| Anonymous | 15,547 | 62.55% | 9,002 | 36.22% | 307 | 1.24% |
| Total | 40,894 | | 33,938 | | 1,475 | |

I then examine the time to update and close an issue, by the gender of the raiser, as shown in Figure 3. Anonymous users tend to update issues quickly, with an average update time of 12.65 days and a closure time of 32.23 days. In contrast, men take longer, averaging 44.00 days to update and 84.04 days to close an issue. Non-binary contributors exhibit the longest times, with an average of 151.56 days to update and 165.58 days to close an issue. Women fall between these groups, taking an average of 20.68 days to update and 54.99 days to close an issue. These differences highlight varying interaction patterns with React's the issue tracking system based on gender identity.

Figure 3: Distribution of Days to Update and Close Issue by Gender

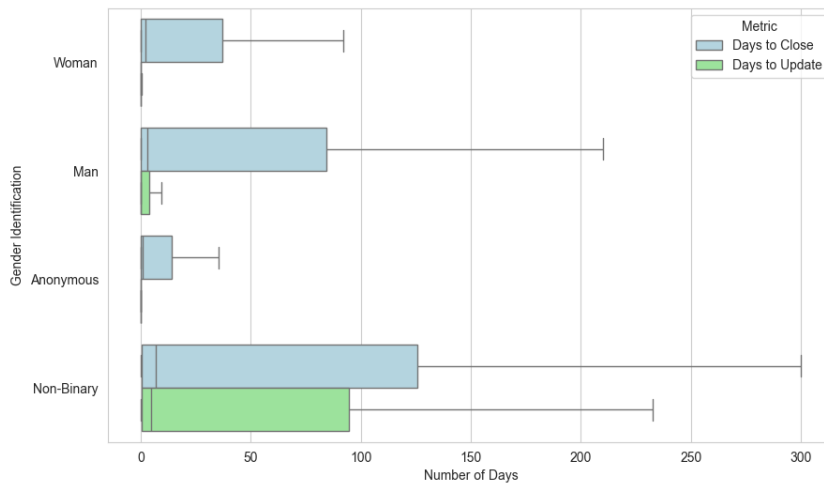

After controlling for the effect of the specific React repository, I find significant differences in the time taken to close an issue between different gender groups (Welch's ANOVA=7.81, p<0.001, df=3259). Therefore, gender is likely a factor influencing how quickly issues are resolved on GitHub.

Again, controlling for repository, Welch's ANOVA revealed significant gender differences in the time to close issues (Welch's ANOVA=272.57, p<0.0001, df=759). The Games-Howell post-hoc test further highlighted differences. Anonymous contributors' close issues significantly faster than both men and women, with a mean difference of 40.39 days compared to men, and 9.23 days compared to women. These differences are statistically significant, with p<0.0001 and p<0.05, respectively. Additionally, men take significantly longer to close issues than women, with an average difference of 31.16 days (p < 0.0001). These results underscore gender and anonymity play a significant role in the efficiency of issue resolution on GitHub.

*4.1.3 Contributor Retention*

Retention of contributors is crucial for the robustness of OSS development, including projects like React, as it ensures continuity, stability, and the preservation of valuable knowledge within the community. Initially I examined at overall patterns of activity over the 11-year history of the React project to see if gendered trends could be observed. Activity here encompasses interactions and contributions represented in Table 2.

Figure 4: Activity on React by Gender by Month

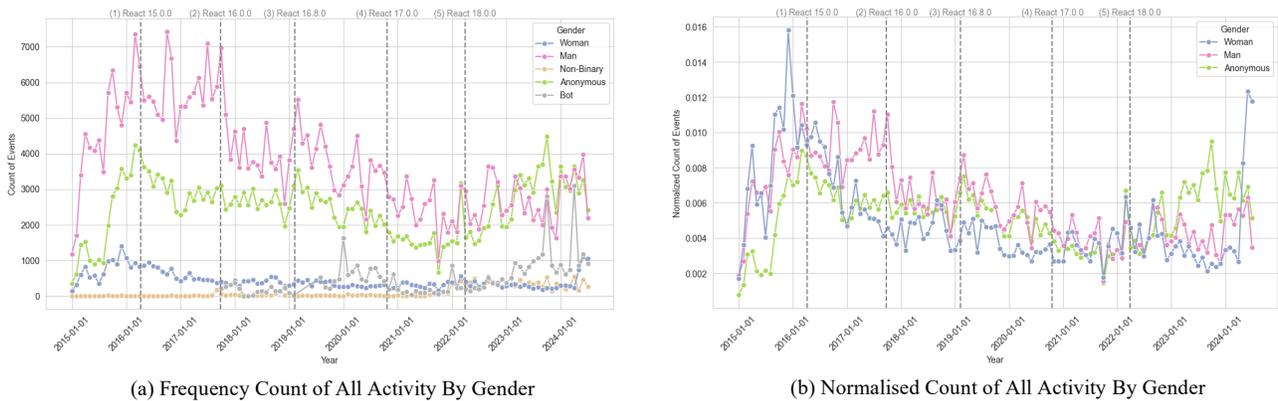

(a) Frequency Count of All Activity By Gender　　　　　　(b) Normalised Count of All Activity By Gender

Figure 4 (b) illustrates that the normalized counts of activity by gender. Normalizing activity counts by the size of each gender group adjusts the raw frequencies to account for different group sizes, allowing for comparisons of activity level that isn't obscured by the size of the group. The graph also highlights milestones that marked significant advancements in React's capabilities and improvements in the developer experience. non-binary contributors are excluded from the normalized counts due to their small sample size, and bots are excluded as they are not relevant to the analysis. For completeness, Figure 5 shows the distribution of users first commit to react over time.

Figure 5: First Contributions to React Over Time by Gender

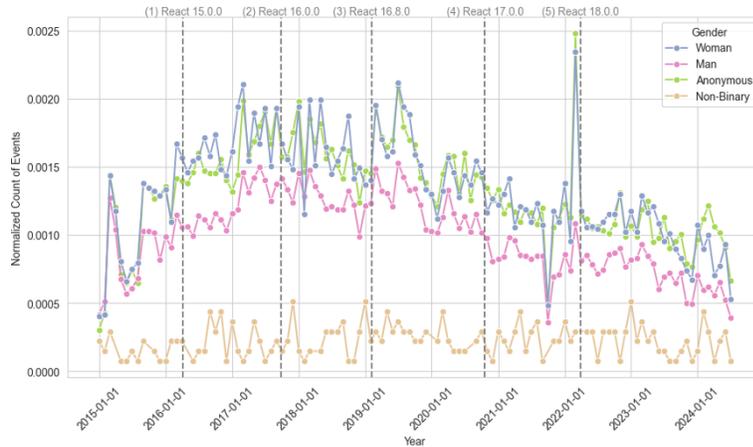

I then examine if gender impacts the retention rate of contributors to React. The retention rate measures the proportion of users from each cohort who remain active in the months following their first contribution, providing insight into how effectively the community retains its contributors over time. I used non-parametric Kaplan-Meier survival curves. Kaplan Meier curves are useful for dealing with time-to-event differences across groups. In terms of retention of users, Kaplan-Meier curves show the cumulative probability of continuing to contribute over a given time. I used two non-parametric statistics to compare Kaplan-Meier survival curves across genders: a log-rank test and a Wilcoxon (Breslow-Gehan) test of the equality of survivor functions.

To summarize, at each event time (measured in months since the first activity), the contribution to the test statistic is calculated as a weighted standardized sum of the difference between the observed and expected number of events in each group. Different tests use different weights. The log-rank test uses a weight of one at all event times and is most suitable when the hazard functions are assumed to be proportional across the groups. Conversely, the Wilcoxon (Breslow-Gehan) test is more appropriate when hazard functions vary in non-proportional ways and when censoring patterns are similar across groups.

Figure 6: Non-Parametric Kaplan-Meier Survival Curves by Gender

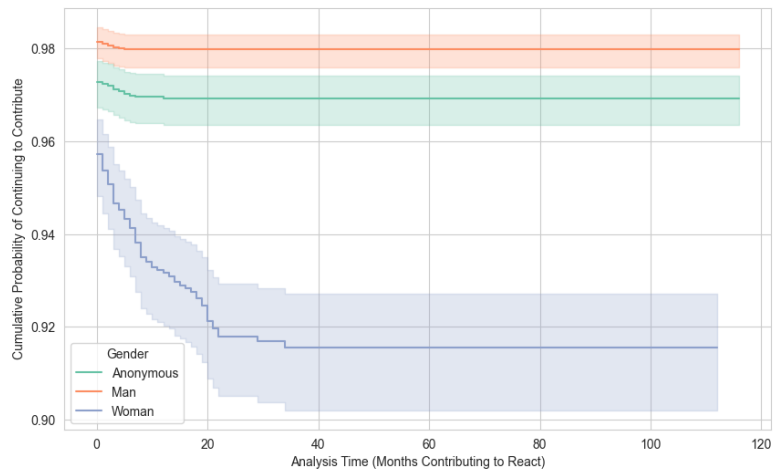

Visual inspection of Figure 6 suggests significant differences in retentions rates between women, men, and anonymous contributors. Women's probability of continuing to contribute declines steeply and substantially in the 20 months from their first contribution. I also conducted a Wilcoxon (Breslow-Gehan) test results revealing significant differences in retention rates across gender groups. For the comparison between men and women, the test statistic is 367.81 ($p < 0.0001$), indicating a highly significant difference. Similarly, the comparison between men and anonymous users yields a test statistic of 227.10 ($p<0.0001$), also signifying a substantial difference. Finally, the comparison between women and anonymous users shows a test statistic of 38.91 ($p<0.0001$), demonstrating a significant difference as well. These results collectively suggest that retention rates vary significantly across different gender groups, with women having the lowest retention rates.

**4.2 Innovation**

The second research question asks: *Does gender impact innovation in the development of React?* To investigate this, I examine if the purpose of contributions vary by gender. A diversity of contributions to a GitHub project indicates innovation by showing that users are enhancing multiple dimensions of the project, bringing varied perspectives and creative solutions to different aspects of its development. I then examine the diversity of contributions over time, focusing on how gender-based contribution patterns shift as key release milestones approach. Since release dates vary slightly across repositories, I focus exclusively on the React.js repository for this analysis.

*4.2.1 Diversity of Contributions*

I ask if innovation in the development of React differences by gender, to examine the potential effect of men's dominance of contributions. Defined in Table 1, I created a typology of PR and commits that represent contributions to the project. There are 120,885 individual PR and commits in the dataset, categorized into: Bug fixing, feature enhancements, documentation, refactoring, dependencies, and testing. The types of contributions are not mutually exclusive, as a PR can both fix a bug and update documentation. Table 5 shows the purpose of contributions to React by gender group.

Table 5: Purpose of Contributions (pull requests and commits) to React by Gender

| **Gender** | **Bug** | **Feature** | **Documentation** | **Refactor** | **Dependency** | **Testing** | **Total** |
|---|---|---|---|---|---|---|---|
| Woman | 3,088 (20.53%) | 3,132 (20.83%) | 2,097 (13.94%) | 2,681 (17.83%) | 2,248 (14.95%) | 1,793 (11.92%) | 15,039 |
| Man | 35,231 (22.41%) | 31,928 (20.31%) | 25,497 (16.22%) | 25,404 (16.16%) | 19,850 (12.62%) | 19,329 (12.29%) | 157,239 |
| Non-Binary | 1,045 (17.31%) | 1,194 (19.78%) | 1,399 (23.18%) | 1,146 (18.99%) | 775 (12.84%) | 477 (7.9%) | 6,036 |
| Anonymous | 22,181 (16.64%) | 27,421 (20.56%) | 25,711 (19.28%) | 30,986 (23.24%) | 13,632 (10.22%) | 13,408 (10.06%) | 133,339 |

Table 5 shows that men and women contribute the most to fixing bugs and enhancing features, whilst non-binary users contribute to documentation and anonymous to refactoring. I then conducted a Chi-squared test to assess if there is a statistically significant difference between genders. I found that: $X^2=4446.23$, $Std(X)^2 = 5.52$, $p<0.001$. Therefore, I conclude that there is a significant difference in patterns of contributions by gender. I then conduct post-hoc analysis through pairwise comparison with a Bonferroni correction. I found no significant difference in comparisons between non-binary users and either man, women, or anonymous users, potentially due to the small sample size. Significant comparisons were found between women and men ($X^2$=152.47, p<0.001), women and anonymous ($X^2$=831.06, p<0.001), and men and anonymous ($X^2$=4135.58, p<0.001).

Taking a broad view of all contributions to React, I find differing patterns of contributions between gender groups, suggesting that the innovation and development processes within the React community are being influenced by a range of diverse perspectives and approaches.

*4.2.2 Releases and New Features*

In the final stage of analysis, I examined the role of gender in innovation by looking at how contribution patterns shift as updated versions of the React library approach release. Initial analysis showed that 54 users are responsible for 491 Release Events, with an average of 9 per user and the most active contributor responsible for 105. The vast majority these contributions were by men, proportionally responsible for 88.80% release events compared to 50.79% of all activity (normalized). In comparison, women were responsible for 6.31% of release events and 7.16% of all activity, non-binary users were 1.43% and 1.54% respectively and anonymous were 3.46% and 34.57%. The low rate of release events by anonymous users is likely because releases are typically handled by Meta (Facebook) employees, who are more inclined to use their real names, as their contributions are part of their professional responsibilities. Nonetheless, men's contributions are disproportionately represented in high-impact events like releases, highlighting a gendered disparity in the roles individuals play in shaping the software's development.

To analyze gender differences in contribution patterns around the release dates for React versions 15.0.0, 16.0.0, 16.8.0, 17.0.0, 18.0.0. I grouped contributions by weeks, categorized by gender and contribution type (Table 1). This allowed me to examine shifts in behavior during specific time windows relative to each release. After establishing a list of release dates and defining time windows to capture periods before, during, and after each release, I filtered the contribution data accordingly. Contributions were then grouped by gender and event category (e.g., bug fixing, feature enhancement) to observe patterns of participation.

Figure 7: Overall Contribution Patterns by Type Approaching React Release

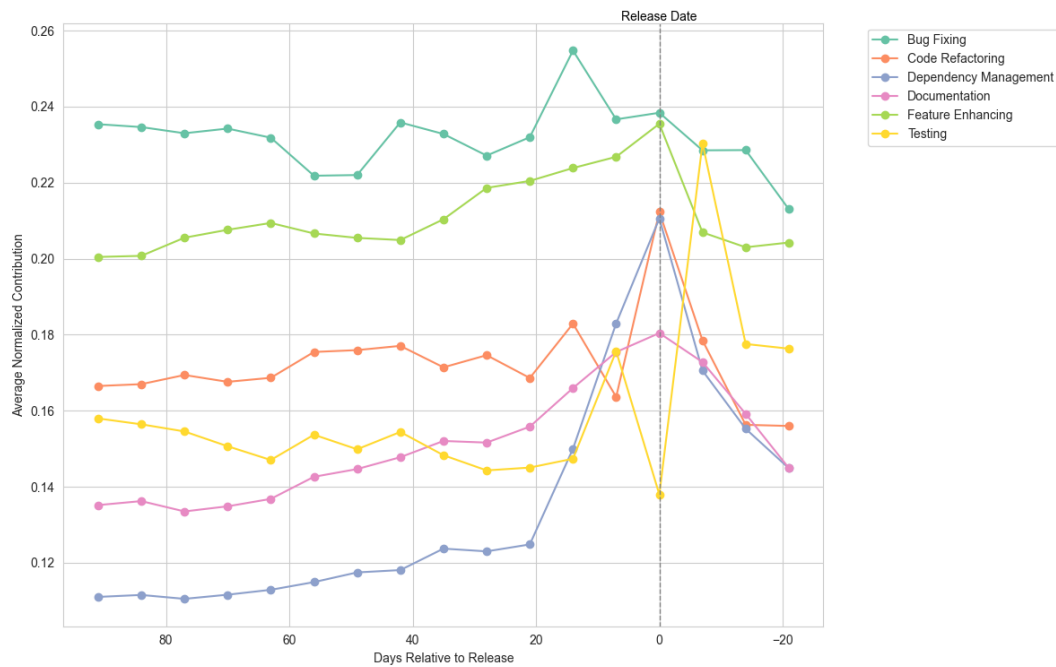

Figure 7 shows overall contribution patterns by type approaching the React release dates identified. Initial inspection reveals that contribution patterns generally began to shift a month before the release. This observation was confirmed with the Wilcoxon signed-rank test, by comparing the frequency of contribution types between pairs of consecutive weekly windows. I found significant differences (p<0.05) in weekly activity by gender from 35 days before the release through 14 days after. These findings suggest that the release cycle significantly impacts the nature and timing of contributions.

I then sought to determine whether these contribution patterns varied by gender. The plots of average contribution patterns by gender approaching and after releases are shown in Figures 8 to 11. I conducted the Kruskal-Wallis which revealed significant gender differences across all six categories (p<0.05) indicating strong evidence of differences in contribution patterns between gender groups. Specifically, accounting for group sizes significant differences were found in bug fixing (H=27.83, p<0.0001), feature enhancing (H=20.19, p<0.0001), testing (H=29.37, p<0.0001), dependency management (H=12.50, p<0.01), documentation (H=21.11, p<0.0001), and code refactoring (H=21.89, p<0.0001).

Figure 8 suggests that men's contribution patterns remain relatively stable as a release approaches, with a decline in code refactoring and an increase in dependency management in the final stages before release. In contrast, Figure 9 shows more noticeable shifts in women's contribution patterns, with significant increases in feature enhancements and documentation as the release date nears. Activities related to bug fixing, dependency management, and testing also rise before release but decline immediately afterward. These increases are expected, as these tasks are essential for ensuring the stability, functionality, and reliability of the software, reflecting the necessary preparation for delivering a high-quality product.

Figure 10 shows that anonymous users focus most on code refactoring and bug fixing prior to release, while also taking a significant role in documentation and testing immediately after the release. Though non-binary users are a smaller group, Figure 11 reveals that their contributions are concentrated around feature enhancements, with declines in documentation and bug fixing leading up to the release, followed by peaks in these activities after the release.

Figure 8: Overall Contribution Patterns for Men by Type Approaching React Release

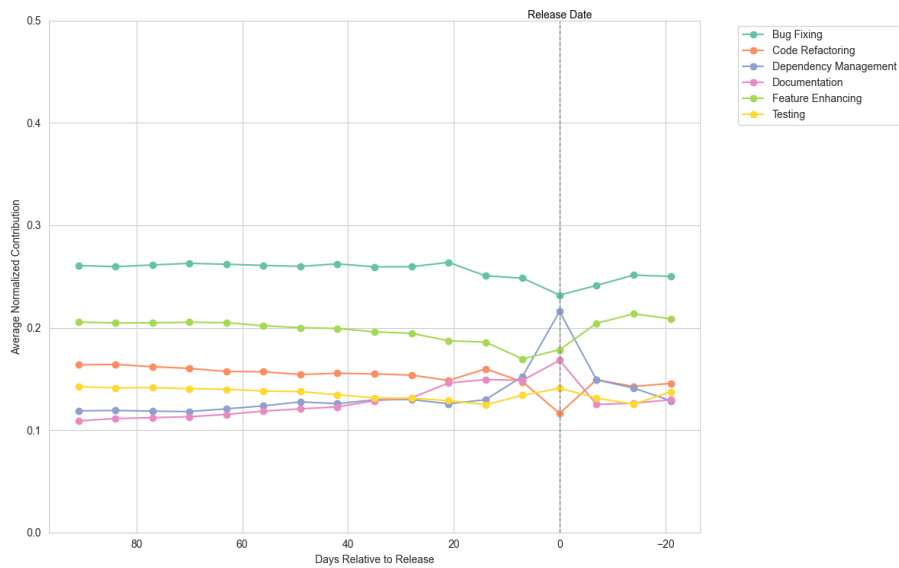

Figure 9: Overall Contribution Patterns for Women by Type Approaching React Release

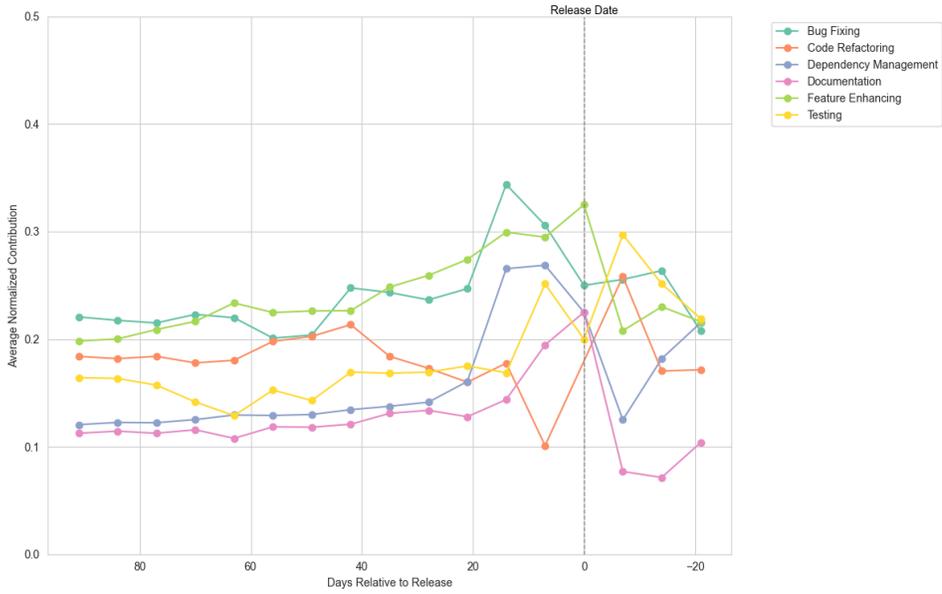

Figure 10: Overall Contribution Patterns for Anonymous by Type Approaching React Release

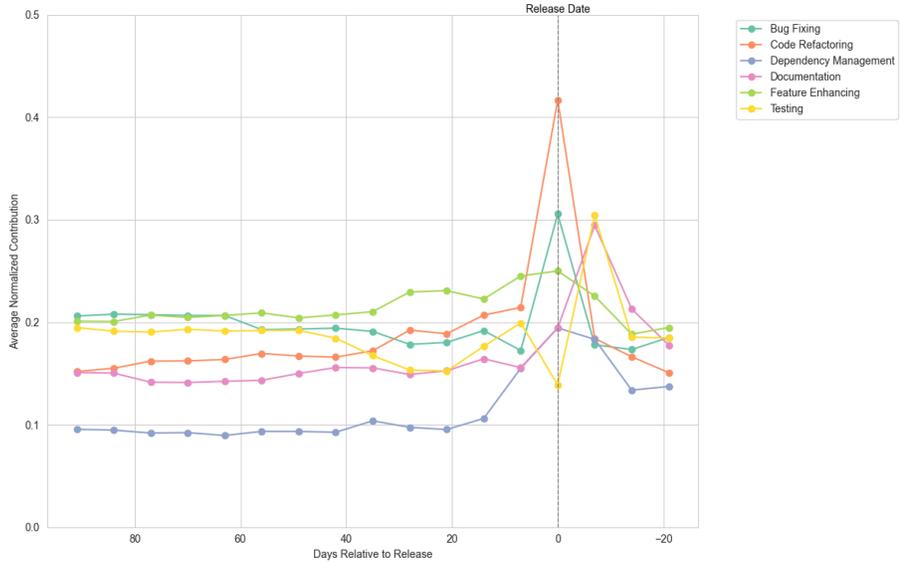

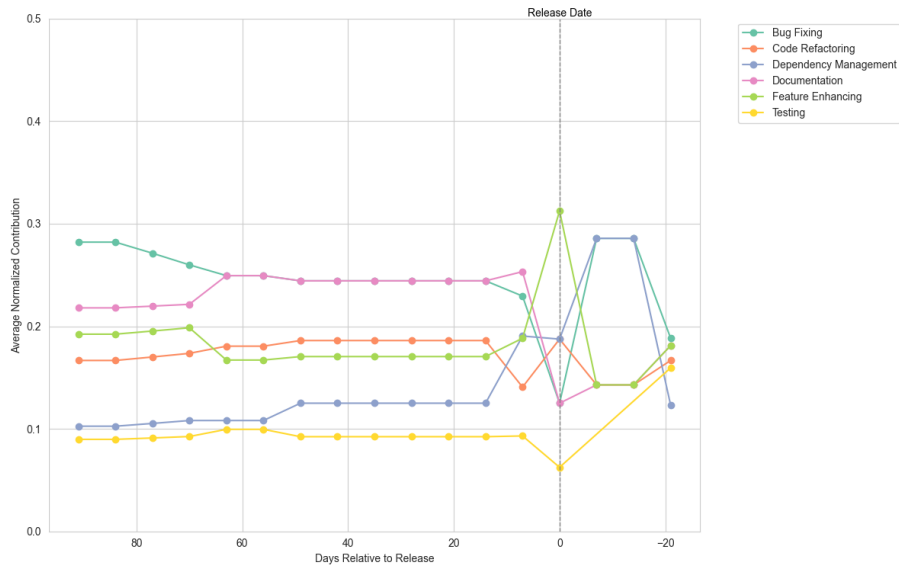

Figure 11: Overall Contribution Patterns for Non-Binary by Type Approaching React Release

Following the Kruskal-Wallis test, a post-hoc analysis using Tukey's HSD test was conducted to identify which specific gender groups differed from each other. In the bug fixing category, the post-hoc analysis revealed that the anonymous group contributed significantly more than both men (Mean Difference=0.0574, p<0.0001) and women (MD=0.043, p<0.0001). However, no significant difference was found between men and women.

For feature enhancing, significant differences were found between women and both the anonymous group (MD= 0.0266, p 0.05) and men (MD=0.0427, p<0.0001), with women contributing more. There was no significant difference between the anonymous group and men.

In the testing category, men contributed significantly more than both the anonymous group (MD=0.0524, p<0.0001) and women (MD=0.0483, p<0.001). No significant difference was found between the anonymous group and women (MD =-0.0041, p>0.05).

In dependency management, the post-hoc analysis revealed that women contributed significantly more than the anonymous group (MD=0.0462, p<0.05). However, no significant differences were found between the anonymous group and men (MD=0.0191, p>0.05) or between men and women (MD=0.0271, p>0.05).

For documentation, the anonymous group contributed significantly less than both men (MD = -0.0386, p<0.05) and women (MD = -0.0414, p<0.001). No significant difference was found between men and women (MD=-0.0028, p>0.05).

Finally, in the code refactoring category, the anonymous group contributed significantly less than men (MD=-0.0352, p<0.05). No significant differences were found between the anonymous group and women (MD=-0.0053, p>0.05) or between men and women (MD=0.0299, p>0.05). In summary, the analysis of contribution patterns approaching and following React releases shows significant gender differences in innovation the development process, with distinct variations in the types of contributions.

## 5 DISCUSSION

### 5.1 Robustness

In addressing my first research question, I found that gender is associated with significant differences in robustness of React's development, though the findings present a complex picture. While there was no gender difference in the time taken for PRs to be merged, a significant difference emerged in the time taken for PRs to be closed, with women's contributions being closed 10.76 days faster than men's. Additionally, women have a lower merge rate than men by over 4%. This suggests that while women's contributions are processed more quickly, they are also less likely to be merged, which may indicate a form of gender bias.

For resolving bugs and issues, women's activities focus more on raising bugs than resolving them whilst men are relatively evenly distributed between raising and resolving. Issues raised by men have the longest time to be both updated and closed, and when men close issues they take 31.61 days longer than women.

While much research has focused on women's retention in OSS projects [17, 31], I tested how the React project compares to existing studies. Ultimately, my findings align with previous research, confirming that women initially participate in higher numbers but experience significantly lower retention rates, likely due to documented biases in inclusion and recognition of their contributions [1, 16]. Notably, anonymous users have significantly higher retention rates than women, suggesting that when a woman's identity is visible through her GitHub username, it negatively impacts her experience and likelihood of continued contribution, aligning with the findings of Terral et a. [29].

The findings have significant implications for the robustness of the React library, particularly regarding the exclusion of women. Robustness in software depends on sustained contributions, which is compromised when women's contributions are undervalued or dismissed. The lower merge rates for women's contributions, despite their quicker processing times, suggest that their work is not being fully integrated into the project, potentially leading to gaps in software quality and resilience. Furthermore, the gender disparity in retention rates reduces the diversity of long-term contributors, weakening the overall robustness of the project.

### 5.2 Innovation

Focusing on the second research question, my results show that gender significantly impacts innovation in the development of React. I found notable differences in the types of contributions made by different gender groups, including between men and women. When analyzing changes in contribution patterns leading up to release milestones, significant gender differences emerged, supporting the theory that gender influences the distribution of tasks that contributors undertake [19, 21]. Specifically, approaching release dates, anonymous users contributed most to bug fixing, men to testing, and women to feature enhancements. Women also contributed significantly more to dependency management compared to anonymous users, but at similar levels to men. Interestingly, men and women contributed equally to documentation, contradicting the perception that women are more involved in organizational or administrative tasks. This suggests that the notion of women's contributions being limited to non-technical roles is more a result of gender bias than an accurate reflection of their actual contributions, in agreement with Nafus [19].

This study's focus on development processes leading up to software releases highlights gender power dynamics, particularly in the unequal distribution of adjustment costs. The study suggests that women often adapt to changing pressures to survive in male-dominated OSS environments. This adaptability underscores the disproportionate burden women face, having to navigate both the demands of the project and gendered expectations. These findings emphasize how power imbalances in tech influence who bears the responsibility of maintaining flexibility as project deadlines approach.

In terms of innovation, the exclusion of women from OSS development poses a serious concern. The results indicate that women contribute significantly to activities associated with innovation, such as feature enhancements and dependency management. By marginalizing women, the OSS community risks missing out on diverse approaches and solutions that could drive creative improvements and advancements in software. Innovation thrives on diverse perspectives and collaborative problem-solving, and when women are excluded or their contributions undervalued, the potential for generating fresh ideas is reduced [21]. Women's exclusion, therefore, not only diminishes equal participation but also curtails the opportunity for OSS projects like React to reach their full innovative potential.

**5.3 Limitations and Future Work**

This study is limited by its focus on a single case study, React. While React is widely used, the findings cannot be generalized to other OSS projects without caution regarding its representativeness. Additionally, the operationalization of robustness and innovation has limitations, as it does not capture the interactive elements of commenting and discussion that occur around contributions. To address this, I incorporated metrics such as time taken to merge and close contributions to include an element of interaction, though analyzing the content of these interactions was beyond the scope of this study. Furthermore, I did not conduct interviews with the developers involved in React, which means my findings lack the depth and richness that qualitative research could provide in understanding the contributors' experiences.

Given that this study provides empirical evidence that gender inclusion impacts OSS development, future research should focus on the purpose of code contributed by different gender groups. As the capabilities of LLMs to analyze and summarize code continue to advance, researchers should explore using these tools to compare gender differences in the purpose and functionality of code contributions. This approach could help test how the inclusion of diverse gender groups might influence the overall functionality and evolution of open-source projects.

**6  CONCLUSIONS AND IMPLICATIONS**

In the introduction I asked *what do we lose by excluding women from software development?* Whilst my results are focused on the case study of React, they do shed some light on this question. I found that woman's exclusion, and a lack of gender diversity more broadly, has measurable implications for the robustness and innovation of software. The lack of full participation by women in software development limits both the diversity of problem-solving approaches and the potential for innovation within projects. This study further emphasizes the gender power dynamics in these environments, suggesting that women often shoulder disproportionate adjustment costs as deadlines approach. They must continuously adapt to both technical demands and gendered expectations in OSS environments, further reinforcing the unequal burden placed on women in the development process.

The OSS community, by excluding women, not only reinforces gender bias but also limits the potential for creating more resilient and innovative software. Beyond providing empirical evidence on how gender impacts both the robustness and innovation of OSS, this project contributes to HCI in three keyways:

- The findings reveal gender bias in OSS, with women having lower merge rates, faster closure of contributions, and steep retention declines within 20 months. This highlights the need for early intervention to improve inclusivity in OSS communities.
- The study introduces a temporal analysis of gendered contribution patterns, showing how contributions shift around key release dates. This offers a new method for analyzing how gender influences development cycles, providing a deeper understanding of how different groups engage with critical phases in software development.

- By linking the exclusion of women to decreased robustness and innovation in software, this research highlights the importance of diversity for both software quality and creative problem-solving, reinforcing diversity as a crucial factor for the success of technological projects in HCI.

Together, these contributions advance HCI by offering new insights into gender dynamics in open-source development, reinforcing the importance of inclusivity for fostering both innovation and robustness in software projects.

## DATASET

[Anonymized for Peer Review]

## FUNDING

[Anonymized for Peer Review]

## ACKNOWLEDGMENTS

[Anonymized for Peer Review]